\documentclass{article}

\usepackage{PRIMEarxiv}

\usepackage[utf8]{inputenc} 
\usepackage[T1]{fontenc}    
\usepackage{hyperref}       
\usepackage{url}            
\usepackage{booktabs}       
\usepackage{amsfonts}       
\usepackage{nicefrac}       
\usepackage{microtype}      
\usepackage{lipsum}
\usepackage{fancyhdr}       
\usepackage{graphicx}       
\graphicspath{{media/}}     
\usepackage{amsmath}
\usepackage{multirow} 

\title{Tag\&Tab: Pretraining Data Detection in Large Language Models Using Keyword-Based Membership Inference Attack}

\author{
    Sagiv Antebi,
    Edan Habler,
    Asaf Shabtai,
    Yuval Elovici\\
    Department of Software and Information Systems Engineering, 
    Ben-Gurion University of the Negev, Israel\\
    \{sagivan,, habler\}@post.bgu.ac.il, \{shabtaia, elovici\}@bgu.ac.il \\ \\
    \url{https://sagivantebi.github.io/tag-tab-site/}
}

\begin{document}
\maketitle

\begin{abstract}
Large language models (LLMs) have become essential tools for digital task assistance.  
Their training relies heavily on the collection of vast amounts of data, which may include copyright-protected or sensitive information. 
Recent studies on detecting pretraining data in LLMs have primarily focused on sentence- or paragraph-level membership inference attacks (MIAs), usually involving probability analysis of the target model's predicted tokens. 
However, these methods often exhibit poor accuracy, failing to account for the semantic importance of textual content and word significance.
To address these shortcomings, we propose Tag\&Tab, a novel approach for detecting data used in LLM pretraining. 
Our method leverages established natural language processing (NLP) techniques to tag keywords in the input text, a process we term \textit{Tagging}.
Then, the LLM is used to obtain probabilities for these keywords and calculate their average log-likelihood to determine input text membership, a process we refer to as \textit{Tabbing}.
Our experiments on four benchmark datasets (BookMIA, MIMIR, PatentMIA, and the Pile) and several open-source LLMs of varying sizes demonstrate an average increase in AUC scores ranging from 5.3\% to 17.6\% over state-of-the-art methods. 
Tag\&Tab not only sets a new standard for data leakage detection in LLMs, but its outstanding performance is a testament to the importance of words in MIAs on LLMs.
\end{abstract}
\section{Introduction}
\label{sec:intro}
The rapid advancement of generative artificial intelligence (GenAI) in recent years has significantly shifted the tech industry's focus toward the development of powerful tools such as large language models (LLMs). 

LLMs are now widely used for tasks such as conversational AI, content generation, and scientific research~\cite{hoang2019efficient,nakano2021webgpt,chatgpt-models,touvron2023LLaMA}. Their adoption reflects a broader shift in AI towards large-scale language understanding.

The widespread use of LLMs has intensified competition to improve model performance, which relies on the collection of vast amounts of data~\cite{wang2023data}.

To achieve these improvements, LLMs are primarily trained on open-source datasets obtained from various sources using methods such as synthetic data generation and web scraping~\cite{nikolenko2021synthetic,khder2021web}.
The type of data collected includes books, code, academic papers, and medical records~\cite{arstechnica_youtube_ai_training,gao2020pile,achiam2023gpt,touvron2023LLaMA}. 
The methods employed for data collection raise significant privacy and ethical concerns~\cite{neel2023privacy,yao2024survey}, primarily regarding the inclusion of personally identifiable information (PII)~\cite{lukas2023analyzing} and copyright-protected content~\cite{rahman2023beyond,wu2024unveiling,arstechnica_youtube_ai_training}.

High-profile lawsuits, such as The New York Times Company vs. OpenAI~\cite{nyt_openai_microsoft_lawsuit}, highlight the need for tools that can detect unauthorized use of data in LLM training~\cite{maini2024llm}.

Membership inference attacks (MIAs) aim to identify whether a given text was part of a model's training data by exploiting behavioral differences in how LLMs process seen versus unseen data (e.g., higher prediction confidence or lower loss)~\cite{hu2022membership,carlini2022membership}.
Existing MIAs face several key limitations. First, most methods rely solely on token-level probabilities, neglecting the semantic importance of words within the broader context~\cite{yeom2018privacy,carlini2021extracting}. Second, their performance varies widely across different models and datasets, often lacking consistent generalization~\cite{duan2024membership,maini2024llm}. Lastly, MIAs are often evaluated on data that is not independently and identically distributed (IID), which can lead to the detection of distribution shifts rather than genuine membership inference, thereby undermining the attacks' reliability~\cite{zhou2023don}.

To address the limitations of existing methods, we introduce Tag\&Tab, a novel approach based on common natural language processing (NLP) methods that is designed to efficiently and effectively detect LLMs' pretraining data. 
Specifically, our method aims to determine whether an LLM was trained on a given text sample, given black-box access to the target LLM (i.e., can only query the model). 

Building on the work of Lukas et al.~\cite{lukas2023analyzing}, who highlighted the role of named entities in PII leakage detection, Tag\&Tab prioritizes informative keywords using entropy-based selection.

Tag\&Tab is designed to address the three key limitations of prior MIA methods. First, rather than relying solely on token-level probabilities, our method introduces semantic awareness by prioritizing meaningful content through keyword selection. Second, our results demonstrate strong generalization across models and datasets, addressing the issue of model inconsistency. Finally, while no MIA is entirely immune to distribution shifts, our focus on rare and informative keywords, rather than shallow statistical artifacts, provides better resilience to distributional variations.

Our method consists of the following steps:
\begin{enumerate}
    \item  Preprocessing - Constructing a word entropy map and filtering certain sentences to ensure optimal keyword selection.
    \item Tagging - Combining (i) Named Entity Recognition (NER)  spans, and (ii) the $K$ words in the sentence with the lowest entropy contribution value. The final set is not constrained to exactly $K$ words.
    \item Tabbing - Passing the entire text to the target LLM and calculating the average log-likelihood of the selected keywords.
    \item Inference - Comparing the average log-likelihood to a threshold to determine the text's membership (i.e., whether it was in the inspected model's training set).
\end{enumerate}

Words with a low entropy contribution are usually either very common or very rare. After constructing the entropy map, we observed that rare content words stand out more in the low-entropy pool, making them more likely to be chosen as keywords.

Our method is based on the intuition that a higher log-likelihood for low-entropy words, together with NER spans, suggests that the model encountered the text during training~\cite{carlini2022quantifying} 
Based on this intuition, we hypothesize that these rare low-entropy words and rare entities are more likely to be memorized by the model and thus serve as effective indicators of the text's membership in the pretraining dataset~\cite{thakkar-etal-2021-understanding,carlini2019secret}.
By selecting a small number of low-entropy keywords and augmenting them with NER spans, our method captures the most informative text elements while minimizing noise from other word probabilities.

We evaluated our method on ten LLMs of varying sizes and four datasets containing nine types of textual data. 
Our results show that Tag\&Tab outperforms state-of-the-art (SOTA) MIAs, achieving an average increase in AUC scores ranging from 5.3\% to 17.6\% compared to the best-performing SOTA method on multiple textual data types.

\noindent The contributions of our paper are as follows: 
\begin{itemize}
    \item We propose Tag\&Tab, a novel approach for the detection of LLMs' pretraining data that focuses on the contextual and semantic relevance of the words in a text and opens the door to additional research on MIAs against LLMs.
    \item To the best of our knowledge, this is the first robust reference-free MIA method to achieve consistently high performance across multiple textual data types and LLMs, outperforming SOTA methods.
    \item Our approach is both resource- and time-efficient. Unlike reference-based attacks that require training a separate model or reference-free methods that depend on auxiliary models (e.g., the Neighbor attack~\cite{mattern2023membership}), Tag\&Tab operates without any additional model training or inference. This minimizes computational overhead and simplifies deployment in real-world scenarios.
\end{itemize}

\begin{figure*}
    \centering
    \includegraphics{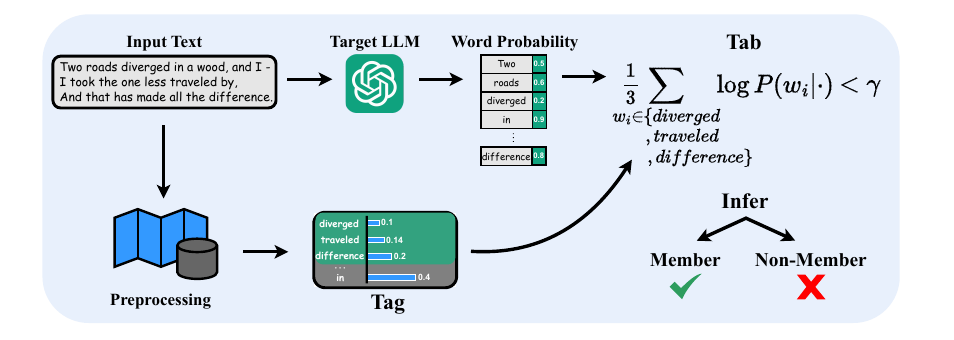}
    \caption{Illustration of the Tag\&Tab method - The process starts by inputting a text (in this example, the input is the conclusion of the well-known poem ``The Road Not Taken"~\cite{frost1916}) in the target LLM to obtain its word probability distribution (word probability). In the tag step, the keywords are selected based on the lowest-entropy words (computed in the preprocessing phase) and NER spans. In the tab step, the log-likelihood of the selected keywords is calculated. Finally, in the infer step, the average log-likelihood of the chosen keywords is compared against a threshold $\gamma$ to determine if the text was part of the target LLM's pretraining data.}
    \label{fig:method_attack}
\end{figure*}

\label{sec:related_work}

\section{Related Work}
\textbf{Membership inference}
(MI)~\cite{shokri2017membership} is a classification task that determines whether a data sample \( x \) was part of a model's training \( D_{\text{train}} \) of a model \( f \). An attacker receives a sample \( x \) and a model \( f \), and applies an attack model \( A \) to classify \( x \) as a member \( x \in D_{\text{train}} \) if \( A(f(x)) = 1 \); otherwise, \( x \) is classified as a non-member \( x \notin D_{\text{train}} \).

\textbf{Large language model membership inference} is a subdomain of membership inference that has gained increasing research attention. Within this subdomain, detecting pretraining data has been the focus of numerous studies exploring different methodologies for determining whether specific texts were included in an LLM's training dataset. Existing MIAs for LLMs fall into two categories: reference-based and reference-free.

\textbf{Reference-based}~\cite{shokri2017membership} methods compare a target model’s outputs to those of reference models, which are typically trained on the same data distribution. One such method is \textit{LiRA}~\cite{carlini2022membership}, which estimates the likelihood ratio of a target example's loss under models trained with and without the example, using Gaussian distributions to simplify the computation.

In contrast, \textbf{reference-free methods} aim to determine membership by applying different probability-based calculations on token predictions.
One such method, the \textit{LOSS Attack}~\cite{yeom2018privacy}, uses model loss values, which in language models correspond to text perplexity.
Perplexity measures how well a probability model predicts a sample and is calculated as the exponentiation of the negative average log-likelihood per token:

{\footnotesize
\[
\text{Perplexity}(P) = \exp\left( - \frac{1}{N} \sum_{i=1}^{N} \log P(t_i \mid t_1, \ldots, t_{i-1}) \right)
\]
}

where \(N\) is the number of tokens, and \(P(t_i \mid t_1, \ldots, t_{i-1})\) is the conditional probability of the \(i\)-th token given its preceding tokens.
 The attack assumes that lower perplexity indicates a text is more familiar to the model, suggesting it was part of the training set. The \textit{Zlib attack}~\cite{carlini2021extracting}, infers membership by calculating the ratio of a text's log-likelihood to its Zlib compression length.
Newer attacks, such as the \textit{Neighbor attack}~\cite{mattern2023membership}, modify selected words in a given text using a different language model to generate 'neighbor' sentences, then compare the original text's perplexity to that of its neighbors. Although the Neighbor attack showed some success, its computational cost is very high compared to other known methods.
More computationally efficient attacks that outperform the Neighbor attack include \textit{Min-K}\%~\cite{shi2023detecting}, and \textit{Min-K\%++}~\cite{zhang2024min}, which focus on the least confident model predictions. Min-K\% calculates the average of the lowest k\% probabilities from the model's output, and Min-K\%++ extends this by normalizing token log probabilities using the mean and variance.
Lastly, two recently published attacks are \textit{ReCaLL}~\cite{xie2024recall}, which measures the relative change in log-likelihood when conditioning the target text on non-member prefixes, and \textit{DC-PDD}~\cite{zhang2024pretraining}, which calibrates token probabilities using divergence from a reference corpus, effectively mitigating the impact of high-frequency tokens.
While each of these attacks has demonstrated success on some datasets and models, their performance remains inconsistent across different studies.
However, recent research performed by Maini et al.~\cite{maini2024llm} showed that aggregating the results of multiple MIAs improves the accuracy of dataset membership inference.
While promising, these findings suggest that the success of this aggregated approach in real-world scenarios depends on improved MIAs, meaning attacks achieving AUC scores above 0.5.

\textbf{Common models} for evaluating MIAs on LLMs include open-source models with known pretraining data.
One such model is LLaMA 1~\cite{touvron2023LLaMA}, created by Meta, which was trained on a mixture of publicly available datasets, including certain subsets of the Pile such as Books3 and Gutenberg, according to Meta’s original paper.

The open-source Pythia model suite~\cite{biderman2023pythia}, which includes eight LLMs ranging from 70M to 12B parameters. These models were trained on data including the Pile dataset, with all models processing the public data in the same order during training.

\section{Method}
\label{sec:method}
We introduce \textit{Tag\&Tab}, a novel resource- and time-efficient method for identifying data used to pretrain LLMs.
\textit{Tag\&Tab} applies common NLP techniques to tag keywords in the pretraining data and predict their membership using the target LLM. 

Our method strategically selects words from the input that should be challenging for the LLM to predict. Successful prediction may indicate that the model previously learned the content during pretraining. 
In Tag\&Tab, words are selected according to their entropy contribution values. Words with a low entropy contribution value tend to be either very common (e.g., 'or,' 'in') or very rare. After constructing an entropy map, we found that the low-entropy pool tends to be dominated by rare content words, which are therefore more likely to be selected as keywords.

Tag\&Tab relies on the hypothesis that while LLMs tend to memorize repeated data, they also memorize rare or unique training data. This aligns with the findings of Carlini et al.~\cite{carlini2022quantifying}, who stated that: "memorization does still happen,
even with just a few duplicates—thus, deduplication will not perfectly prevent leakage." This insight was further validated by Lukas et al.~\cite{lukas2023analyzing}, who found that PIIs tokens, which are distinctive rare tokens, are effective in revealing memorized content.
Thus, an LLM is likely to assign a higher probability to these low-entropy words when they appear in familiar contexts it has seen before, compared to the same low-entropy words in novel or unfamiliar contexts.
For example, in the sentence "Astronomers detected light from the edge of the universe emitted by a quasar," the word "quasar" is a low-entropy word. If this sentence appeared in the training data, the model will assign a higher probability to "quasar" in this context than if it had not seen the sentence before.

Tag\&Tab selects $K$ words with the lowest entropy values in a sentence, which are referred to as \textit{keywords}. By selecting a small number of informative keywords, it aims to capture the semantic importance of the pretraining content. This approach hinges on a hypothesis that only $K$ selected keywords are needed to accurately predict the membership of the entire input text in the pretraining data, minimizing noise from other words in the input text~\cite{shi2023detecting}. 

Tag\&Tab operates under black-box constraints, meaning that we can observe token probabilities from a given input but lack access to the model’s weights, which is standard practice in MIAs~\cite{truex2019demystifying,hu2022membership,mattern2023membership,zhang2024min}.

The \textit{Tag\&Tab} method consists of four stages, which are illustrated in Figure~\ref{fig:method_attack}: 
\begin{enumerate}
    \item{\textbf{Preprocessing}} - 
    First, the \textbf{word entropy map} is constructed using the Python package \textit{wordfreq}~\cite{robyn_speer_2022_7199437}, which provides frequency estimates for words in a specified language. The entropy contribution value for each word is calculated using the formula:
    \[
    H(w_i) = - p(w_i) \cdot \log_2 p(w_i)
    \]
    This value is high for mid-frequency words and low for both very common and very rare words. 
    In this stage, the text is also split into individual sentences, using segmentation tools (e.g., the NLTK package~\cite{bird2009natural}). To avoid selecting less informative keywords due to insufficient sentence length, sentences with fewer than a specified number of words are filtered out.
    \item{\textbf{Tagging}} - From each sentence \(S \) in the text file \( T \), the method first takes the \( K \) lowest-entropy words according to the entropy map. We then identify named-entity spans (e.g., people, organizations, locations) using spaCy~\cite{spacy2}. The final keyword set is the union of the low-entropy words and the NER spans. Because this is a union, the set may be larger than \( K \), but in practice it is often exactly \( K \), because NER spans often overlap with low‑entropy tokens. For clarity, we continue to denote the size of this final set as \(K\), referring to the number of keywords after the union.

    \item {\textbf{Tabbing}} - This stage mimics the auto-completion feature found in interfaces like a command line, where it predicts and fills in the rest of the command based on the context of the preceding input. 
    Using the target model $M$, the method computes the log-likelihood of the entire text, then focuses on the log-likelihood of the previously identified keywords. For each sentence \( S \in T\) consisting of \( n \) words \( w_1, w_2, \ldots, w_n \), where each word \( w_i \) is decomposed into tokens, denoted as \( w_i = t_{i_1}, t_{i_2}, \ldots, t_{i_m} \), token \( t_{i_j} \), given its preceding tokens, is calculated as \( \log p_M(t_{i_j} | t_{i_1}, \ldots, t_{i_{j-1}}) \).
    We define the log-likelihood of a word \( w_i \) using the log-likelihood of its first token \( t_{i_1} \) given its preceding tokens, expressed as \( \log p_M(t_{i_1} | t_{1_1}, t_{1_2}, \ldots, t_{i-1_j}) \). 
As a result, we obtain:
\[
  \log p_M(w_i \mid \cdot) 
  = \log p_M(t_{i_1} \mid  \cdot)
\]

The method selects the \( K \) keywords from \( S \)  and computes the average log-likelihood of the keywords:

    {\footnotesize
    \[
    \text{Keywords' Prob}(S) = \frac{1}{K} \sum_{w_i \in 
    \text{Keywords}(S)} \log p_M(w_i | \cdot) 
    \]
    }

    \item {\textbf{Inferring}} - In this stage, the method calculates the average probability of the keywords across all sentences in text $T$ and compares it to a predetermined threshold $\gamma$ to determine membership.
\end{enumerate}

\section{Evaluation}
\label{sec:evaluation}

This section presents a detailed evaluation of Tag\&Tab's effectiveness. The experiments were conducted on a NVIDIA RTX 6000 GPU, running for nearly three days on all models and datasets. We used the default parameter settings of widely adopted libraries, including spaCy and NLTK.

\subsection{Model Comparison}
To compare \textit{Tag\&Tab} with other reference-free baseline detection methods, we examined various open-source LLMs, including LLaMA 1 (7B, 13B, 30B)~\cite{touvron2023LLaMA}, Pythia (160M, 1.4B, 2.8B, 6.9B, 12B)~\cite{biderman2023pythia}, and Qwen1.5-14B~\cite{qwen2024qwen15}. Additionally, we included GPT-3.5 Turbo\footnote{https://platform.openai.com/docs/models/gpt-3-5-turbo} (trained on data up to September 2021\footnote{https://learn.microsoft.com/en-us/azure/ai-services/openai/concepts/models}), given partial knowledge of its training on known books, as discussed in previous studies~\cite{shi2023detecting,chang2023speak}. Our black-box assumption still holds because the OpenAI API exposes token-level log probabilities for this model. LLaMA 1 and Pythia are well-suited for MIA evaluation due to their transparency regarding pretraining datasets, unlike newer models such as LLaMA 2 and 3, which lack such transparency.

\subsection{Dataset Comparison}
The experiments were conducted on the BookMIA\footnote{\url{www.huggingface.co/datasets/swj0419/BookMIA}}~\cite{shi2023detecting}, Pile\footnote{\url{www.huggingface.co/datasets/monology/pile-uncopyrighted/viewer/default/validation}}~\cite{gao2020pile}, and MIMIR\footnote{\url{www.huggingface.co/datasets/iamgroot42/mimir}}~\cite{duan2024membership} datasets, meeting the requirement that MIA evaluation datasets should be as comprehensive and diverse as possible~\cite{duan2024membership}, covering various types of text while maintaining a consistent distribution between training and test sets.

The Pile is a collection of diverse texts of different types from a wide range of sources, designed to train and evaluate LLMs using open-source data. In our experiments, for each domain we used 10,000 samples for training and an additional 10,000 samples from the same domain for testing. BookMIA is a dataset consisting of 100 books, with 100 text chunks of 512 words extracted from each, totaling 10,000 samples. The dataset is evenly split into member and non-member sets. The member set includes 5,000 samples from 50 books from the Books3 dataset~\cite{reisner2023books3}, a collection of ebooks, most of which were published between 2000 and 2020. The non-member set comprises 5,000 samples from 50 new books with first editions
in 2023 that could not have been in the training data of GPT-3.5 Turbo~\cite{chang2023speak}.
BookMIA is designed to evaluate MIAs by contrasting model behavior on seen and unseen content. 
For our evaluation on Pythia and LLaMA 1, we leveraged the fact that their training corpora, the Pile and Books3 respectively, have significant overlap with the Gutenberg dataset. The BookMIA member set, sourced from Books3, contains 50 books, 34 of which are also present in Gutenberg. To construct a high-confidence member set, we focused our evaluation on this specific overlap, filtering out the 16 member books from BookMIA that are not part of the Gutenberg corpus. This ensures that our test set consists of books with the strongest possible evidence of inclusion in the pretraining data, providing a more controlled experimental setup. We note that cutoff-based datasets such as BookMIA have been critiqued as unsuitable for strict benchmarking of MIAs due to temporal distribution shifts~\cite{das2025blind, maini2024llm}. Accordingly, we do not treat BookMIA as a primary benchmark but rather as an illustrative case study simulating a practical audit scenario.
 MIMIR is a dataset built from the Pile, designed to evaluate memorization in LLMs. 
 MIMIR is a post-processed subset of the Pile, intended to reduce distributional artifacts such as temporal drift and stylistic shifts. It contains data known to have been used in training all Pythia model sizes, offering a unified benchmark for assessing membership inference. In our evaluation, we utilized around 2,000 samples per domain (approximately 1,000 member and 1,000 non-member samples).

The PatentMIA\footnote{\url{https://github.com/zhang-wei-chao/DC-PDD/tree/main}}~\cite{zhang2024pretraining} corpus was used in a dedicated non-Latin evaluation on the Chinese language, using the Qwen1.5-14B model. The setup and results are discussed in Appendix~\ref{sec:patentmia}.

It is important to note that we opted not to evaluate our method on the WikiMIA dataset,\footnote{www.huggingface.co/datasets/swj0419/WikiMIA}~\cite{shi2023detecting} as recent publications (e.g., \cite{maini2024llm}) questioned the reliability of the data due to temporal shifts in writing styles and an insufficient number of samples.

\subsection{Evaluation Approach}
To assess our method's performance, we followed a systematic process that begins with the input dataset. Each text file in the dataset is processed, with every document truncated to a maximum of 2,048 tokens to ensure a consistent input size. Sentences with fewer than seven words are excluded. We identify and save the K keywords for every sentence.
We documented the outcomes of selecting 1 to 10 entropy keywords per sentence (before adding the NER words).

After selecting the K keywords, the entire text is processed by the target model, which outputs probability distributions for each token. 
Then we average the log-likelihoods of all tokens in the keywords, conditioned on their preceding tokens, assessing the model's familiarity with the entire keyword. 

Following Carlini et al.'s evaluation process~\cite{carlini2022membership}, we set a threshold to assess attack performance, focusing on TPR at low FPR, denoted as T@F.

We also report the area under the ROC curve (AUC score) to provide a clearer measure of detection performance. The AUC score quantifies the overall performance of a classification method by considering TPRs and FPRs at all classification thresholds. Since AUC offers a comprehensive, threshold-independent evaluation metric, we do not need to determine a specific threshold \(\gamma\) for our method.

To simulate a real-world application, Appendix~\ref{sec:threshold-calibration} details how we pick a working threshold~\(\gamma\) when only non–member data from the same domain are available. The appendix further shows that applying this book-derived threshold to a different domain (mathematics) degrades performance, so \(\gamma\) must be recalibrated for every model–dataset pair.

\begin{table*}[ht]
    \centering
    \scriptsize
    \setlength{\tabcolsep}{2.5pt}
    \renewcommand{\arraystretch}{1.2}
    \caption{Comparison of AUC results for Tag\&Tab and baseline methods on the MIMIR and Pile benchmarks. The upper table presents the best results from Tag\&Tab and baseline methods across four MIMIR datasets, while the lower table shows the best results for three Pile datasets. The best results for each dataset and model size are highlighted in bold, and the second-best AUC is underscored.}
    \begin{tabular}{lcccccccccccccccccccc}
        \toprule
        \multirow{2}{*}{Method} & \multicolumn{5}{c}{DM Mathematics} & \multicolumn{5}{c}{GitHub} & \multicolumn{5}{c}{Pile CC} & \multicolumn{5}{c}{C4} \\
        \cmidrule(lr){2-6} \cmidrule(lr){7-11} \cmidrule(lr){12-16} \cmidrule(lr){17-21}
        & 160M & 1.4B & 2.8B & 6.9B & 12B & 160M & 1.4B & 2.8B & 6.9B & 12B & 160M & 1.4B & 2.8B & 6.9B & 12B & 160M & 1.4B & 2.8B & 6.9B & 12B \\
        \midrule
        Loss & 0.85 & 0.76 & 0.84 & 0.68 & 0.86 & 0.80 & 0.85 & 0.86 & 0.88 & 0.88 & 0.53 & 0.54 & 0.54 & 0.55 & 0.55 & 0.50 & 0.51 & 0.51 & 0.51 & \textbf{0.51} \\
        Zlib & 0.68 & 0.59 & 0.66 & 0.55 & 0.69 & \underline{0.84} & \underline{0.88} & \underline{0.89} & \underline{0.90} & \underline{0.90} & 0.51 & 0.53 & 0.53 & 0.54 & 0.54 & 0.51 & 0.51 & 0.51 & 0.51 & \textbf{0.51} \\
        Min-20\% Prob & 0.61 & 0.53 & 0.70 & 0.50 & 0.82 & 0.80 & 0.85 & 0.86 & 0.88 & 0.88 & 0.52 & 0.53 & 0.54 & 0.55 & 0.55 & 0.51 & 0.51 & 0.51 & 0.51 & 0.50 \\
        Max-20\% Prob & 0.63 & 0.67 & 0.61 & 0.58 & 0.51 & 0.78 & 0.85 & 0.85 & 0.87 & 0.86 & 0.52 & 0.53 & 0.53 & 0.53 & 0.54 & 0.51 & 0.50 & 0.50 & 0.50 & 0.50 \\
        Min‑K\%++ (20\%) & 0.81 & 0.79 & 0.66 & 0.81 & 0.73 & 0.57 & 0.57 & 0.61 & 0.63 & 0.66 & 0.51 & 0.50 & 0.52 & 0.53 & 0.53 & 0.52 & 0.51 & 0.51 & 0.50 & 0.50 \\
        ReCaLL & 0.80 & 0.73 & 0.78 & 0.64 & 0.86 & 0.79 & 0.76 & 0.74 & 0.71 & 0.72 & 0.53 & 0.54 & 0.54 & 0.55 & 0.55 & 0.51 & 0.51 & 0.51 & 0.51 & \textbf{0.51} \\
        DC-PDD & 0.90 & 0.86 & 0.86 & 0.85 & 0.86 & \textbf{0.87} & \textbf{0.91} & \textbf{0.92} & \textbf{0.93} & \textbf{0.93} & \underline{0.54} & 0.55 & \textbf{0.56} & \textbf{0.57} & \textbf{0.57} & 0.51 & 0.51 & 0.51 & 0.51 & \textbf{0.51} \\
        \midrule

        Ours (Tag\&Tab K=4) & \textbf{0.96} & \textbf{0.96} & \textbf{0.96} & \textbf{0.95} & \textbf{0.95} & 0.78 & 0.82 & 0.83 & 0.84 & 0.85 & \underline{0.54} & \textbf{0.56} & \textbf{0.56} & \textbf{0.57} & \textbf{0.57} & \textbf{0.53} & \textbf{0.52} & \textbf{0.52} & \textbf{0.52} & \textbf{0.51} \\
        Ours (Tag\&Tab K=10) & \underline{0.92} & \underline{0.92} & \underline{0.93} & \underline{0.92} & \textbf{0.95} & 0.79 & 0.83 & 0.84 & 0.85 & 0.86 & \textbf{0.55} & \textbf{0.56} & \textbf{0.56} & \textbf{0.57} & \underline{0.56} & \textbf{0.53} & \textbf{0.52} & \textbf{0.52} & \textbf{0.52} & \textbf{0.51} \\
        \midrule
        \multirow{2}{*}{Method} & \multicolumn{5}{c}{Ubuntu IRC} & \multicolumn{5}{c}{Gutenberg} & \multicolumn{5}{c}{EuroParl} & \multicolumn{5}{c}{Average} \\
        \cmidrule(lr){2-6} \cmidrule(lr){7-11} \cmidrule(lr){12-16} \cmidrule(lr){17-21}
        & 160M & 1.4B & 2.8B & 6.9B & 12B & 160M & 1.4B & 2.8B & 6.9B & 12B & 160M & 1.4B & 2.8B & 6.9B & 12B & 160M & 1.4B & 2.8B & 6.9B & 12B \\
\midrule
Loss & 0.63 & 0.59 & 0.60 & 0.58 & 0.58 & 0.53 & 0.53 & 0.53 & 0.53 & 0.53 & 0.52 & 0.52 & 0.50 & 0.52 & 0.51 & 0.67 & 0.67 & 0.69 & 0.66 & 0.70 \\

Zlib & 0.52 & 0.52 & 0.53 & 0.54 & 0.54 & 0.53 & 0.60 & 0.53 & 0.53 & 0.53 & 0.51 & 0.51 & 0.50 & 0.51 & 0.51 & 0.63 & 0.63 & 0.65 & 0.62 & 0.66 \\

Min-20\% Prob & 0.58 & 0.57 & 0.52 & 0.51 & 0.52 & 0.53 & 0.53 & 0.53 & 0.53 & 0.60 & 0.53 & \textbf{0.54} & 0.52 & 0.50 & 0.51 & 0.61 & 0.61 & 0.65 & 0.61 & 0.69 \\

Max-20\% Prob & \underline{0.69} & \textbf{0.69} & \textbf{0.71} & \textbf{0.68} & \textbf{0.67} & \textbf{0.67} & \underline{0.73} & 0.60 & \underline{0.67} & 0.67 & 0.53 & \textbf{0.54} & \textbf{0.55} & 0.53 & 0.55 & 0.61 & 0.64 & 0.62 & 0.62 & 0.60 \\

Min‑K\%++ (20\%) & 0.52 & 0.51 & 0.52 & 0.54 & 0.61 & \textbf{0.67} & 0.60 & 0.60 & 0.60 & 0.60 & 0.54 & 0.53 & 0.51 & 0.51 & 0.51 & 0.60 & 0.59 & 0.57 & 0.62 & 0.61 \\

ReCaLL & \textbf{0.72} & 0.64 & \underline{0.69} & 0.64 & 0.60 & 0.53 & \textbf{0.80} & \textbf{0.67} & \textbf{0.73} & \textbf{0.80} & 0.51 & 0.51 & 0.51 & 0.55 & \textbf{0.57} & 0.67 & 0.64 & 0.65 & 0.62 & 0.68 \\

DC-PDD & 0.58 & 0.53 & 0.53 & 0.53 & 0.53 & 0.53 & 0.60 & 0.60 & 0.53 & 0.53 & 0.51 & 0.52 & 0.50 & 0.51 & 0.54 & \textbf{0.70} & 0.70 & 0.72 & 0.71 & 0.70 \\
\midrule

Ours (Tag\&Tab K=4) & 0.64 & \underline{0.65} & 0.64 & \underline{0.66} & \underline{0.64} & \textbf{0.67} & 0.67 & \textbf{0.67} & \underline{0.67} & 0.67 & \underline{0.55} & \textbf{0.54} & \textbf{0.55} & \textbf{0.54} & \underline{0.56} & \textbf{0.70} & \textbf{0.72} & \textbf{0.73} & \textbf{0.72} & \textbf{0.73} \\

Ours (Tag\&Tab K=10) & 0.61 & 0.63 & 0.62 & 0.61 & 0.62 & 0.60 & 0.67 & \textbf{0.67} & \underline{0.67} & 0.67 & \textbf{0.56} & \textbf{0.54} & \textbf{0.55} & \textbf{0.54} & 0.55 & \textbf{0.70} & \underline{0.71} & \textbf{0.73} & \textbf{0.72} & \underline{0.72} \\

\bottomrule
    \end{tabular}
    \label{tab:results_mimir_pile}

\end{table*}

\section{Results}
\label{sec:results}
This section presents the results of two case studies using Tag\&Tab, each examining a different aspect of pretraining data detection in LLMs. Each case study is evaluated using the following metrics: AUC and TPR at a low FPR of 5\% (T@F=5\%).

\begin{figure}
    \centering
\includegraphics{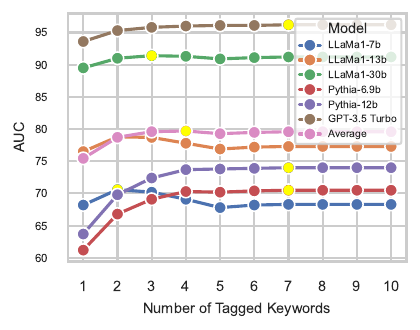}
\caption{AUC scores as a function of the number of tagged keywords for the examined models on the BookMIA dataset. Yellow points indicate optimal performance: 2-3 keywords for LLaMA 1, 7 for Pythia and GPT-3.5 Turbo, and 4 on average (Average).}
\label{fig:best_k}
\end{figure}

Throughout this section, we report the results for Tag\&Tab using \(K=4\) and \(K=10\), as these values yielded the most consistent and high-performing outcomes across models and datasets. This choice is supported by the results presented in Figure~\ref{fig:best_k}, which shows that Tag\&Tab performs robustly for \(K\) values between 4 and 10 for most of the models, with minimal performance variation—making the method resilient to non-optimal keyword selections.

Case Study 1 serves as the primary benchmark evaluation, while Case Study 2 is included to illustrate a more practical, engineered audit scenario closer to real-world use.

The reported results are based on a single run, as we observed minimal variation across multiple runs.

\subsection{Case Study 1: Detecting Various Types of Pretraining Data in LLMs}
This case study evaluates Tag\&Tab's effectiveness and robustness in detecting different types of pretraining data in LLMs. 
We evaluate our method on various sizes of the Pythia model and compare its effectiveness against baseline attacks on seven types of text in the Pile and MIMIR datasets.
Table~\ref{tab:results_mimir_pile} summarizes the results obtained when targeting five Pythia model sizes ranging from 160M to 12B parameters, tested with two configurations of tagged keywords: 4 and 10.

\noindent The main findings from these results are as follows:
\begin{itemize}
\item When averaging performance across all datasets, Tag\&Tab ($K=4$) outperforms all baseline methods over all the models, establishing itself as the most effective approach overall. Tag\&Tab ($K=10$) ranks as the second-best method, demonstrating strong performance but falling short of the results achieved with $K=4$.

\item Notably, Tag\&Tab ($K=4$) achieves either the best or second-best results for the majority of textual data types examined. Even when it does not come in first or second place, its performance remains competitive, serving as a robust alternative to the leading method.

\item While Tag\&Tab ($K=10$) selects more keywords, increasing the number of probabilities considered for text membership inference, its results are consistently lower than those of Tag\&Tab when $K=4$. This supports the hypothesis that selecting a smaller number of keywords allows the method to extract noise-free information from the model. 
\end{itemize}

We also observe that certain formal texts, such as mathematical proofs, may contain fewer named entities or conventional keywords. However, they often feature domain-specific terminology or symbolic expressions that carry strong membership signals. This is evidenced by our results on the DM Mathematics subset (Table~\ref{tab:results_mimir_pile}), where Tag\&Tab maintained SOTA performance, achieving an AUC between 0.95 and 0.96.

Despite outperforming the baseline MIAs, the AUC achieved by Tag\&Tab is relatively low in certain cases, hovering around 0.55.  However, recent research by Maini et al.~\cite{maini2024llm} showed that aggregating multiple MIAs improves dataset membership inference accuracy, emphasizing the need for better attacks that achieve AUC over 0.5 for improved aggregated attack performance. 
Tag\&Tab meets this criterion, making it a valuable component in an ensemble of MIAs, as we further demonstrate in Appendix~\ref{appendix:ensemble}.

\begin{table*}[ht]
\centering
\small
\setlength{\tabcolsep}{2pt}
\begin{tabular}{l cc cc cc cc cc cc cc cc}
\toprule
\multirow{2}{*}{\textbf{Method}} & \multicolumn{2}{c}{\textbf{LLaMA-7b}} & \multicolumn{2}{c}{\textbf{LLaMA-13b}} & \multicolumn{2}{c}{\textbf{LLaMA-30b}} & \multicolumn{2}{c}{\textbf{Pythia-6.9b}} & \multicolumn{2}{c}{\textbf{Pythia-12b}} & \multicolumn{2}{c}{\textbf{GPT-3.5}} & \multicolumn{2}{c}{\textbf{Average}} \\ 
\cmidrule(lr){2-3} \cmidrule(lr){4-5} \cmidrule(lr){6-7} \cmidrule(lr){8-9} \cmidrule(lr){10-11} \cmidrule(lr){12-13} \cmidrule(lr){14-15}
 & \textit{AUC} & \textit{T@F5} & \textit{AUC} & \textit{T@F5} & \textit{AUC} & \textit{T@F5} & \textit{AUC} & \textit{T@F5} & \textit{AUC} & \textit{T@F5} & \textit{AUC} & \textit{T@F5} & \textit{AUC} & \textit{T@F5} \\ 
\midrule
Neighbor & 0.65 & \underline{0.27} & 0.71 & 0.38 & 0.90 & 0.73 & 0.65 & 0.26 & 0.71 & 0.36 & 0.96 & 0.88 & 0.76 & 0.48 \\
Loss & 0.59 & 0.25 & 0.70 & 0.43 & 0.89 & 0.74 & 0.62 & 0.24 & 0.69 & 0.32 & \textbf{0.97} & \underline{0.90} & 0.74 & 0.48 \\ 
Zlib & 0.53 & 0.22 & 0.67 & 0.42 & 0.89 & 0.74 & 0.55 & 0.19 & 0.61 & 0.25 & 0.96 & 0.88 & 0.70 & 0.45 \\ 
Min-20.0\% Prob & 0.61 & 0.24 & 0.70 & 0.42 & 0.87 & 0.70 & 0.65 & 0.25 & 0.70 & 0.34 & 0.95 & 0.86 & 0.75 & 0.47 \\ 
Min‑K\%++ (20\%) & 0.60 & 0.23 & 0.68 & 0.38 & 0.78 & 0.60 & 0.59 & 0.20 & 0.56 & 0.20 & 0.95 & 0.86 & 0.69 & 0.41 \\ 
Max-20.0\% Prob & 0.51 & 0.15 & 0.66 & 0.34 & 0.87 & 0.69 & 0.51 & 0.13 & 0.59 & 0.20 & 0.96 & \textbf{0.91} & 0.68 & 0.40 \\ 
ReCaLL & 0.58 & 0.22 & 0.70 & 0.42 & 0.84 & 0.64 & 0.66 & 0.29 & 0.72 & \underline{0.37} & 0.74 & 0.50 & 0.70 & 0.41 \\ 
DC-PDD & 0.61 & 0.27 & 0.71 & \underline{0.47} & 0.88 & \textbf{0.77} & 0.68 & \textbf{0.34} & 0.74 & \textbf{0.44} & 0.95 & 0.89  & 0.76 & \textbf{0.53} \\ 
\midrule
Ours (Tag\&Tab K=4) & \textbf{0.69} & \textbf{0.28} & \textbf{0.78} & \textbf{0.48} & \textbf{0.91} & 0.76 & \textbf{0.72} & \underline{0.30} & \underline{0.75} & 0.36 & \textbf{0.97} & \underline{0.90} & \textbf{0.80} & \underline{0.51} \\ 
Ours (Tag\&Tab K=10) & \underline{0.67} & 0.26 & \underline{0.77} & 0.46 & \textbf{0.91} & \textbf{0.77} & \textbf{0.72} & \underline{0.30} & \textbf{0.76} & 0.36 & 0.96 & 0.87 & \textbf{0.80} & 0.50 \\
\bottomrule
\end{tabular}
\caption{Detection of data from the BookMIA dataset used in pretraining seven models using Tag\&Tab and six baseline MIAs, evaluated in terms of AUC and T@F=5\%. All results are reported as decimal fractions. The last two rows compare the Tag\&Tab method when selecting four and ten keywords. The best results are bolded, and second-best are underscored.}
\label{tab:main_results}
\end{table*}

\subsection{Case Study 2: Detecting Specific Pretraining Data in LLMs}
This case study focuses on the precision in detecting specific pretraining data in LLMs.
We designed a targeted attack to infer whether copyrighted data was part of the model's pretraining. Unlike Case Study 1, the member and non-member data come from different sources.
Using the BookMIA dataset, we simulate partial knowledge of a model's pretraining data to infer specific text files suspected of being included in the target model's training set.
For validation, we selected non-members from books published after the target model's release thus ensuring they were not part of its pretraining data.

To determine the optimal number of keywords to select, we evaluate the results by selecting between 1 and 10 keywords from each sentence
The results shown in Figure~\ref{fig:best_k} demonstrate that the optimal number of keywords required to ensure effective detection depends on the model architecture. As can be seen, for different sizes of the LLaMA 1 model, the optimal number of keywords ranged from 2 to 3, while for the Pythia models and GPT-3.5 Turbo, the optimal number of tagged keywords was 7.
The best results across all models were achieved with $K=4$, yielding an average AUC score of 0.8.

Table~\ref{tab:main_results} summarizes the results of Tag\&Tab and reference-free baseline attacks.

\noindent The main findings from these results are as follows:

\begin{itemize}
\item Tag\&Tab outperforms all other attacks in terms of the AUC, with an average improvement of 5.3\%–17.6\% over baseline methods when $K=4$.

\item Tag\&Tab ($K=4$) outperforms other attacks in terms of the T@F=5\% on the LLaMA-7B and LLaMA-13B models. However, for most other models, the DC-PDD attack performs better, with Tag\&Tab consistently ranking second.

\item As the size of the examined LLM increases, the AUC scores of the MIAs also increase due to the model's memorization capacity \cite{carlini2022quantifying}. This can be seen in the results for Tag\&Tab which achieved very high AUC scores: (1) 0.91 on LLaMA-30b compared to 0.69 on LLaMA-7b, (2) 0.75 on Pythia-12b compared to 0.72 on Pythia-6.9b, and (3) 0.97 on GPT-3.5 Turbo, the largest model tested.
\end{itemize}

To better understand our method's performance, in Appendix~\ref{sec:results-k-selection}, we examine the impact of our method's tagging stage by comparing the selection of the lowest $K$ entropy words with random token selection, observing that prioritizing the lowest $K$ entropy words significantly enhances performance across all models, resulting in superior AUC scores.

Finally, Appendix~\ref{sec:ablation-study} reports the results obtained on additional experiments on our tagging choices. Using only named‑entity tokens or only low‑entropy tokens improves on the baselines, but combining the two as done in Tag\&Tab produces the best results. We also experimented with ranking tokens based on TF-IDF rather than entropy, which resulted in noticeably poorer performance.

\section{Conclusion}
We present Tag\&Tab, a novel black-box method for detecting pretraining data in LLMs. By focusing on the semantic and contextual relevance of words, the method enhances detection capabilities. Tag\&Tab outperforms SOTA attacks and consistently achieves high performance across diverse textual data types.
Our comprehensive evaluation spans nine textual data types from four datasets (the Pile, MIMIR, BookMIA, and PatentMIA) and ten LLMs of varying sizes and architectures (LLaMA 1-7b, 13b, 30b, Pythia-160m, 1.4b, 2.8b, 6.9b, 12b, GPT-3.5 Turbo, Qwen1.5-14b).
Our study confirms that the selection of low-entropy contribution keywords, augmented with NER spans, improves membership inference attack results, further validating our approach. Future work could extend Tag\&Tab by considering keyword context and placement within documents. Additionally, developing new MIAs that leverage advanced NLP techniques to assess word significance could further improve the detection of pretraining data in LLMs.

\bibliographystyle{unsrt}  
\bibliography{references}  

\appendix

\section{Appendix}
Our code is available in our GitHub repository.~\footnote{https://github.com/sagivantebi/Tag-Tab}

\subsection{Impact of Our Word Selection Process}
\label{sec:results-k-selection}
\begin{figure*}
    \centering
    \includegraphics{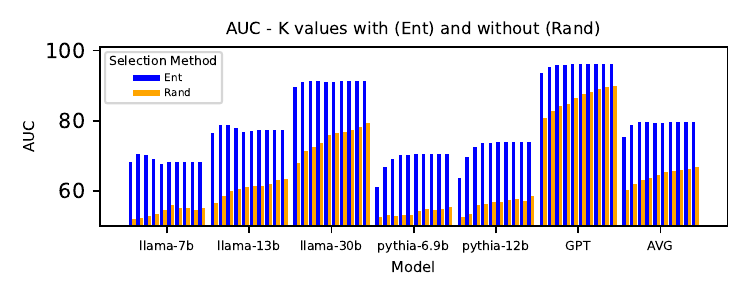}
\caption{
Comparison of Tag\&Tab’s Tabbing performance with our original Tagging keyword selection against Tabbing performance with random word selection for different $K$ values. For each model, we report the AUC scores obtained when selecting 1-10 keywords per text using the Tag\&Tab entropy-based with NER spans process (blue) and a random word selection method (orange). Each bar represents the AUC achieved for a specific $K$ value. Results are presented for all evaluated models on the BookMIA dataset as well as the overall average (AVG). The plot highlights the consistent advantage of low-entropy keyword selection across varying $K$ values and model sizes.
}
    \label{fig:tag_vs_random}
\end{figure*}

To evaluate the impact of our word selection process, Tagging, we compared it against a random selection of words using the same Tabbing algorithm. Tagging selects the lowest $K$ entropy words, augmented with NER spans,  while the random baseline selects words without regard to entropy. Figure~\ref{fig:tag_vs_random} illustrates the results of this comparison.
The dataset used in this evaluation was BookMIA. For each model, we report the AUC scores obtained when selecting between one and ten keywords per text using the Tag\&Tab entropy-based method (blue) and a random word selection (orange). The results show that Tag\&Tab's original Tagging process improves performance across all models. Tag\&Tab's original Tagging process achieved an average AUC of 0.8, compared to the average AUC of 0.64 obtained by randomly selecting $K$ words. This demonstrates the Tag method's effectiveness in enhancing model performance by focusing on rare words.

\subsection{Ablation Study: Alternative Word Selection Strategies}
\label{sec:ablation-study}

To analyze the contribution of different word selection strategies, we conducted an ablation study comparing several variants of our method on the BookMIA dataset. Specifically, we evaluated the following alternatives to the Tagging process used by the Tag\&Tab method:

\begin{itemize}
    \item \textbf{Entropy-Only}: selecting the four lowest entropy tokens per sentence.
    \item \textbf{NER-Only}: selecting the first four named entities identified via spaCy per sentence.
    \item \textbf{TF-IDF}: selecting the top four tokens with the highest TF-IDF scores per sentence.
\end{itemize}

We evaluated these variants on all of the models except GPT-3.5 Turbo.

The results, summarized in Table~\ref{tab:ablation_results}, show that both the \textit{Entropy-Only} and \textit{NER-Only} variants achieved competitive performance, with AUC values slightly below those achieved when the original Tagging process is used by the Tag\&Tab method. The \textit{TF-IDF} variant performed notably worse, confirming that low-entropy and named-entity tokens are more effective indicators of membership when used together. These findings validate our design choice of combining both strategies for optimal pretraining data detection.

\begin{table*}[h]
\centering
\caption{AUC scores of Tag\&Tab and the evaluated variants on the BookMIA dataset, all with $K=4$ selected keywords.}
\label{tab:ablation_results}
\begin{tabular}{lccccc}
\toprule
\textbf{Method} & \textbf{LLaMA-7B} & \textbf{LLaMA-13B} & \textbf{LLaMA-30B} & \textbf{Pythia-6.9B} & \textbf{Pythia-12B} \\
\midrule
Tag\&Tab & \textbf{0.69} & \textbf{0.78} & \textbf{0.91} & \textbf{0.72} & \textbf{0.75} \\
Entropy-Only & 0.67 & 0.76 & 0.89 & 0.70 & 0.73 \\
NER-Only  & 0.62 & 0.72 & 0.85 & 0.68 & 0.73 \\
TF-IDF         & 0.60 & 0.68 & 0.81 & 0.65 & 0.67 \\
\bottomrule
\end{tabular}
\end{table*}

\begin{figure}[h]
    \centering
    \includegraphics[width=1.1\linewidth]{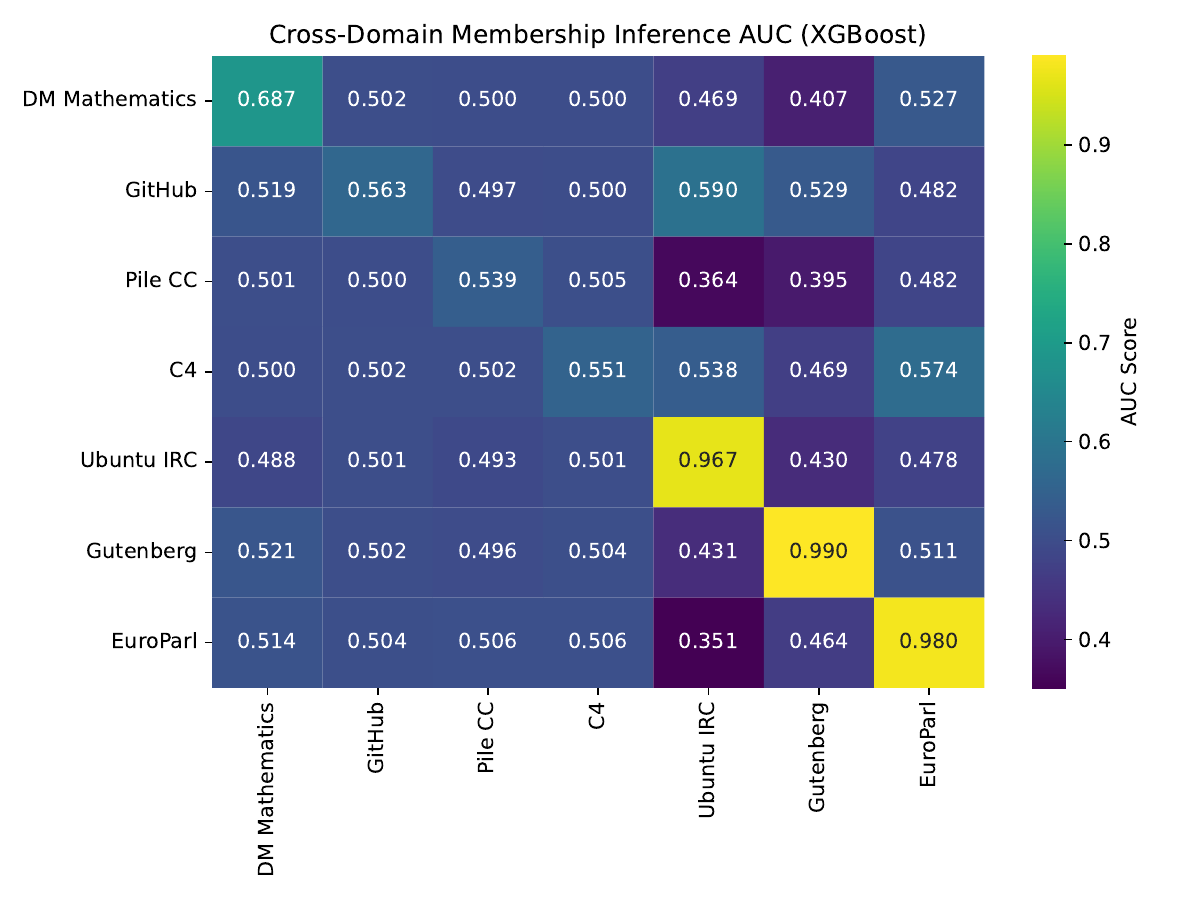}
    \caption{Cross-domain membership inference AUC scores using XGBoost classifiers (threshold’s transferability). 
    Each row corresponds to a classifier trained on the domain listed, and each column indicates the evaluation domain. 
    Diagonal values represent in-domain evaluation, while off-diagonal entries show cross-domain transfer performance.}
    \label{fig:crossdomain_heatmap}
\end{figure}

\subsection{Threshold Calibration}
\label{sec:threshold-calibration}

\subsubsection{Calibrating \(\gamma\) with No Member Labels}
While in a realistic “zero-knowledge” setting, we assume no access to the model’s training data, we can still collect data that were not seen during training. This includes texts published after the model’s release or synthetically generated samples.  
To calibrate the decision threshold~\(\gamma\), we run Tag\&Tab on a set of known non‑member samples and select \(\gamma\) based on the upper tail of Tag\&Tab's scores distribution (e.g., the 95\textsuperscript{th} percentile).  
This threshold is then used to flag samples that appear “member-like” relative to the known non‑members.  
Using BookMIA with LLaMA‑30B and \(K{=}4\), this approach yielded \(\gamma \approx 0.0392\), resulting in an AUC of 0.85 on the test split, a modest drop from the 0.91 AUC achieved when member labels were available.
Although the performance was slightly poorer, the method still achieved strong results, demonstrating its potential for practical real-world applications even under limited knowledge conditions.

\subsubsection{Domain Mismatch Test}
To examine the threshold's transferability, we applied the same \(\gamma=0.0392\) to a math corpus from MIMIR (DM Mathematics), using the same model, LLaMA‑30B.  
Performance dropped sharply (AUC $<0.60$), showing that Tag\&Tab's scores distribution differs across domains.  
We saw the same effect when switching models and concluded that each model–dataset pair needs its own calibrated threshold.

\subsubsection{Cross-Domain Generalization}
To further investigate domain transferability, we extended our thresholding approach beyond a fixed scalar \(\gamma\).  
Instead of using a single cutoff, we trained a supervised XGBoost classifier on Tag\&Tab scores extracted from each of seven domains (DM Mathematics, GitHub, Pile CC, C4, Ubuntu IRC, Gutenberg, and EuroParl).  
In this setup, the XGBoost model functions as a learned, domain-specific thresholding mechanism: it establishes a more flexible decision boundary in score space that generalizes the role of \(\gamma\).  

Each trained classifier was then evaluated across all domains, yielding the 7×7 AUC matrix shown in Figure~\ref{fig:crossdomain_heatmap}.  
This experiment therefore examines whether thresholds calibrated in one domain can transfer effectively to other domains.

The results reveal a clear trend: classifiers achieve the highest AUC when evaluated on the same domain they were trained on (diagonal entries), but their performance drops sharply when applied to different domains. For example, the classifier trained on Ubuntu IRC data achieves an AUC of 0.967 when tested on Ubuntu IRC data but falls below 0.50 on most other domains. Similarly, classifiers trained on Gutenberg (0.990) and EuroParl (0.980) achieve near-perfect within-domain performance but transfer poorly elsewhere.

These findings confirm that Tag\&Tab’s inference thresholds are highly domain-specific.

\subsubsection{Recommendation}
Thresholds should be recalibrated whenever the target \emph{model} or \emph{data domain} changes.  
A small, trustworthy non‑member dev set from the intended domain is sufficient; no labeled members are required.

\subsection{Generalization to Chinese Texts}
\label{sec:patentmia}
To examine whether Tag\&Tab can generalize to a non-Latin language with a fundamentally different structure, we evaluated it on a Chinese text using the PatentMIA dataset~\cite{zhang2024pretraining}, which contains patents sourced from Google Patents~\cite{googlepatents2006}.
 This evaluation was performed using the Qwen1.5-14B model,\footnote{https://huggingface.co/Qwen/Qwen1.5-14B} an open-source LLM developed by Alibaba Cloud, optimized for Chinese and multilingual understanding~\cite{qwen2024qwen15}. Tag\&Tab still achieved a good score (AUC = 0.6), but it did not surpass the strongest baseline (DC‑PDD, 0.69 AUC). We attribute this gap to structural differences between Chinese and English, such as the absence of explicit word boundaries and different entropy distributions, which diminish the effectiveness of our current keyword–entropy heuristic.

\subsection{Robustness to Adversarial and Distributional Perturbations}
\label{appendix:robustness}

To evaluate Tag\&Tab's robustness against minor textual modifications, we conducted an experiment where 2–5 words per sample in the BookMIA dataset were replaced with suitable synonyms. These changes preserved the original meaning while altering the lexical form, simulating both adversarial-style perturbations and natural distribution shifts.

We evaluated the impact of these modifications on the LLaMA-7B, LLaMA-13B, LLaMA-30B, Pythia-6.9B, and Pythia-12B models. In all cases, we used Tag\&Tab with $K=4$ keywords.

The results, presented in Table~\ref{tab:robustness}, show that Tag\&Tab exhibits only a minor performance drop of 1–2\% in the AUC across all models. These findings confirm that the method remains effective even when the exact data distribution is unknown, demonstrating resilience to small-scale semantic-preserving shifts.

\begin{table}[h]
\centering
\caption{Robustness of Tag\&Tab (K=4) under synonym-based perturbations on the BookMIA dataset.}
\label{tab:robustness}
\begin{tabular}{lcc}
\toprule
\textbf{Model} & \textbf{Original AUC} & \textbf{Perturbed AUC} \\
\midrule
LLaMA-7B     & 0.69 & 0.68 \\
LLaMA-13B    & 0.78 & 0.76 \\
LLaMA-30B    & 0.91 & 0.89 \\
Pythia-6.9B  & 0.72 & 0.71 \\
Pythia-12B   & 0.75 & 0.73 \\
\bottomrule
\end{tabular}
\end{table}

\subsection{Ensemble Membership Inference Attacks}
\label{appendix:ensemble}

While Tag\&Tab achieves strong standalone results, recent work has shown that ensemble methods, combining multiple MIA scores, often outperform individual attacks~\cite{maini2024llm}. To evaluate whether Tag\&Tab score contributes complementary information, we trained a simple XGBoost classifier on features extracted from different attack methods. Experiments were conducted on the BookMIA dataset using the LLaMA-30b model. The feature sets used to train the classifiers included: (i) Tag\&Tab scores only, (ii) an ensemble of existing methods' scores (PPL, Zlib, Min-20\%), and (iii) the ensemble augmented with Tag\&Tab scores.

The results, summarized in Table~\ref{tab:ensemble}, demonstrate that Tag\&Tab substantially improves ensemble performance. While the ensemble of other methods achieved an AUC of 0.94, adding Tag\&Tab raised the AUC to 0.97. These findings confirm that Tag\&Tab captures complementary signals not fully exploited by existing approaches, strengthening its value as a component in broader attack pipelines. 

\begin{table}[h]
\centering
\caption{Ensemble membership inference attack performance on the BookMIA dataset and LLaMA-30b.}
\label{tab:ensemble}
\begin{tabular}{lc}
\toprule
\textbf{Feature Set} & \textbf{AUC} \\
\midrule
Tag\&Tab only & 0.91 \\
Ensemble (PPL, Zlib, Min-20\%) & 0.94 \\
Ensemble + Tag\&Tab & \textbf{0.97} \\
\bottomrule
\end{tabular}
\end{table}

\end{document}